# Quantum identity authentication protocol via perturbed probability amplitudes

Xiangqun Fu*, Zheng Han, Hongwei Li, Chun Zhou, Shuo Zhang, Jianhong Shi, Tan Li, Wansu Bao

Henan Key Laboratory of Quantum Information and Cryptography, Information Engineering University, Zhengzhou, 450001, China

**Abstract**: Quantum identity authentication is an effective solution for establishing mutual trust between communicating parties in the post-quantum era. To address the risk of partial secret information leakage under special measurements, this paper designs a quantum identity authentication protocol based on perturbed probability amplitudes. The protocol first proposes a method of encoding a random sequence onto the probability amplitudes of quantum entangled states, forming perturbed probability amplitudes, thereby reducing the involvement of shared information between the sender and receiver. Furthermore, it presents a method for generating *n*-bit random authentication information based on the *n*-bit pre-shared information, enhancing the randomness of the authentication information relative to an attacker. The proposed protocol can resist impersonation attack, CNOT attack, and entanglement measurement attack, and ensures backward security. Under equivalent security conditions, the amount of shared information is 1/*n* that in Chen et al.'s scheme and half that in Dutta et al.'s scheme. And each bit of the secret message has no risk of leakage under special measurements. It can also be extended to a bi-directional quantum identity authentication protocol.

## 1 Introduction

In today's highly interconnected digital era, ensuring the true identity of communication participants is the foundation for building a secure information system. Identity authentication, as the first line of defense for information security, has the core task of verifying whether a user claiming a certain identity is indeed a legitimate entity, thereby preventing malicious attackers from stealing sensitive information or compromising communication security by impersonating legitimate users (e.g., through man-in-the-middle attacks, replay attacks, etc.).

* Corresponding author (Xiangqun Fu, email: fuxiangqun@126.com)

Traditional identity authentication protocols primarily rely on computationally hard mathematical problems (such as large integer factorization and discrete logarithms), whose security is based on the assumption of limited computational power. However, with the rapid development of quantum computing technology, quantum algorithms such as Shor's algorithm have proven capable of solving these problems in polynomial time, posing an unprecedented security threat to traditional authentication protocols. Therefore, it is urgent to develop novel authentication mechanisms.

In 1995, Crépeau et al. proposed the first quantum identity authentication (QIA) protocol[1], whose security is based on the principles of quantum mechanics. Unlike classical authentication, QIA utilizes quantum states (such as single photons or entangled photon pairs) as information carriers or authentication tokens, such that any eavesdropping attempt will inevitably perturb the quantum system and thus be detected by the legitimate users. Since the late 1990s, researchers have proposed various QIA protocols, and their technical approaches can be mainly classified into the following categories.

**(1) Entanglement-based schemes**: These schemes utilize the strong correlation properties of multi-particle entangled states such as EPR pairs[2-6], GHZ states[7-10] and two-mode squeezed state[11] to achieve identity authentication. Such schemes generally offer high security and can realize bidirectional or simultaneous multi-party authentication, but they impose high technical requirements for the preparation and maintenance of quantum resources.

**(2) Single-photon-based schemes**[12-16]: These schemes use only single-photon states as quantum resources, drawing on or improving the framework of quantum key distribution protocols such as BB84. Such schemes have attracted widespread attention due to their better compatibility with current technology and higher efficiency.

**(3) Hybrid and specialized schemes**: These schemes combine technologies such as quantum key distribution[17-18], quantum secure direct communication[19-20], quantum teleportation[4,21], quantum secret sharing[22-25], and quantum computing tasks[26-31] to derive specialized protocols such as controlled authentication, deniable authentication, and semi-quantum authentication, in order to meet the requirements of specific application scenarios.

From the perspective of shared information, existing QIA protocols can be mainly classified into the following categories:

**(1) Shared classical key**[6,15,28,32-35]: In shared classical key QIA protocols, the communicating parties pre-share a predetermined classical bit string to authenticate each other's identity. The core idea of this scheme is that both parties utilize the pre-shared classical key string to achieve the dual functions-identity verification and eavesdropping detection-by correlating key bits with the positions and measurement bases of the transmitted particles.

**(2) Shared entangled state**[36-38]: This is a more profound form of "shared information". The two parties pre-share one or more entangled states (such as EPR pairs). The inherent quantum correlation of these entangled states themselves constitutes a unique form of "shared information" that cannot be classically replicated. The authentication process verifies whether this quantum correlation remains intact through operations such as Bell state measurements. Due to the quantum no-cloning theorem, an attacker cannot copy this "token". Furthermore, well-designed protocols can keep the entangled state unchanged or reset it after successful authentication, thereby achieving the renewability of the authentication token and solving the problem of the shared key being exhausted and requiring frequent updates.

**(3) Shared identity identifier and hash function**[3,10,39]: This type mainly appears in multi-party authentication protocols. The user and trusted third party pre-share a unique identity number (ID) and a secret hash function. The trusted third party uses this information to operate on the quantum states, and the user verifies their identity through the correlation of the measurement results.

In fact, the greatest advantage of existing QIA protocols over classical identity authentication protocols is that there is no need to send the message code to the verifier, only the authentication code needs to be sent. However, the authentication code is often encoded into a quantum state. If a third party selects an appropriate measurement basis to measure the authentication code, the third party can deterministically obtain the private information. For example, in Ref.[33], the authentication code generated based on shared information is $|0\rangle$ or $|-\rangle$. If the attacker measures in the horizontal basis and obtains $|+\rangle$, then the sent authentication code is $|0\rangle$, meaning the exact shared information can be obtained. Therefore, encoding different private information into different quantum states poses security risks, and it is necessary to further select an appropriate encoding method to enhance the security of QIA.

In this paper, we design a quantum identity authentication via perturbed probability amplitude. The secret message is encoded onto the probability amplitudes of the entangled state, and combining perturbing the probability amplitudes, it ensures that each bit of the secret message has no risk of leakage under special measurements.

The paper is arranged as follows. In Sec. 2, we design the quantum identity authentication protocol via perturbed probability amplitude. In Sec. 3, we give the security analyze and the comparison of the new protocol. Finally, the conclusion is presented.

## 2 Quantum identity authentication protocol via perturbed probability amplitude

**(1) Protocol conditions**

Alice and Bob share a key $K=\left(k_0,k_1,\cdots,k_{n-1}\right)$ , where $k_i$ = 0 or 1 and $i=0,1,\cdots,n-1$.

**(2) Perturbation**

*Step* 1: Alice prepares the particles $a_i$ and $b_i$ in the quantum entangled state $\left|\phi_i\right\rangle=\frac{1}{\sqrt{2}}\left(\left|0_{a_i}0_{b_i}\right\rangle+\left|1_{a_i}1_{b_i}\right\rangle\right)$, where $i=0,1,\cdots,n-1$. And $\left|\Phi\right\rangle=\otimes_{i=1}^{n}\left|\phi_i\right\rangle$ .

*Step* 2: Alice chooses a random sequence $r=\left(r_0,r_1,\cdots,r_{n-1}\right)$, and applies the single-qubit rotation $R_y\left(r_i\theta\right)=\begin{bmatrix}\cos\frac{r_i\theta}{2} & -\sin\frac{r_i\theta}{2}\\ \sin\frac{r_i\theta}{2} & \cos\frac{r_i\theta}{2}\end{bmatrix}$ on the particle $a_i$ , where $r_i=0$ *or* 1 and $\theta=\frac{\pi}{4}$.

$$\left|\phi_i\right\rangle\xrightarrow{R_y\left(r_i\theta\right)}\frac{1}{\sqrt{2}}\left(R_y\left(r_i\theta\right)\left|0_{a_i}\right\rangle\left|0_{b_i}\right\rangle+R_y\left(r_i\theta\right)\left|1_{a_i}\right\rangle\left|1_{b_i}\right\rangle\right)=\left|\phi_i'\right\rangle$$

*Step* 3: Alice sends the particles $(a_0,a_1,\cdots,a_{n-1})$ to Bob.

**(3) Encryption**

*Step* 4: Bob receives $(a_0,a_1,\cdots,a_{n-1})$ and performs the following rotation transformation on each $a_i$ according to the shared key:

i) If $k_i=0$,

$$\left|\phi_i'\right\rangle\xrightarrow{R_y((k_i\oplus k_{(i+1)\bmod n}+1)\theta)}$$
$$\frac{1}{\sqrt{2}}\left(R_y((k_i\oplus k_{(i+1)\bmod n}+1+r_i)\theta)\left|0_{a_i}\right\rangle\left|0_{b_i}\right\rangle+R_y((k_i\oplus k_{(i+1)\bmod n}+1+r)\theta)\left|1_{a_i}\right\rangle\left|1_{b_i}\right\rangle\right)=\left|\phi_i''\right\rangle$$

ii) If $k_i=1$,

$$\left|\phi_i'\right\rangle \xrightarrow{R_y((k_i\oplus k_{(i+2)\bmod n}+1)\theta)}$$

$$\frac{1}{\sqrt{2}}\left(R_y((k_i\oplus k_{(i+2)\bmod n}+1+r_i)\theta)\left|0_{a_i}\right\rangle\left|0_{b_i}\right\rangle+R_y((k_i\oplus k_{(i+2)\bmod n}+1+r_i)\theta)\left|1_{a_i}\right\rangle\left|1_{b_i}\right\rangle\right)=\left|\phi_i''\right\rangle$$

*Step* 5: Bob sends $(a_0,a_1,\cdots,a_{n-1})$ back to Alice.

**(4) Decryption and authentication**

*Step* 6: Alice receives $(a_0,a_1,\cdots,a_{n-1})$ and performs the following rotation transformation on each $a_i$ according to the shared key:

i) If $k_i=0$,

$$\left|\phi_i''\right\rangle \xrightarrow{R_y(-(k_i\oplus k_{(i+1)\bmod n}+1+r_i)\theta)} \frac{1}{\sqrt{2}}\left(\left|0_{a_i}0_{b_i}\right\rangle+\left|1_{a_i}1_{b_i}\right\rangle\right)=\left|\phi_i\right\rangle$$

ii) If $k_i=1$,

$$\left|\phi_i''\right\rangle \xrightarrow{R_y(-(k_i\oplus k_{(i+2)\bmod n}+1+r_i)\theta)} \frac{1}{\sqrt{2}}\left(\left|0_{a_i}0_{b_i}\right\rangle+\left|1_{a_i}1_{b_i}\right\rangle\right)=\left|\phi_i\right\rangle$$

*Step* 7: Alice measures $a_i$ and $b_i$ in the Bell basis, and obtains $c_i$ ($i=0,1,\cdots,n-1$). If $c_0=c_1=\cdots=c_{i-1}=0$, Alice passes Bob's identity authentication; otherwise, it is not passed.

**NOTE**: The Bell basis $\frac{1}{\sqrt{2}}(\left|00\right\rangle+\left|11\right\rangle)$, $\frac{1}{\sqrt{2}}(\left|00\right\rangle-\left|11\right\rangle)$, $\frac{1}{\sqrt{2}}(\left|01\right\rangle+\left|10\right\rangle)$ and $\frac{1}{\sqrt{2}}(\left|01\right\rangle-\left|10\right\rangle)$ are denoted as 0, 1, 2 and 3 respectively.

**The correctness of decryption and authentication**

In *Step* 6, if $k_i=0$,

$$\left|\phi_i''\right\rangle=\frac{1}{\sqrt{2}}\left(R_y((k_i\oplus k_{(i+1)\bmod n}+1+r_i)\theta)\left|0_{a_i}\right\rangle\left|0_{b_i}\right\rangle+R_y((k_i\oplus k_{(i+1)\bmod n}+1+r)\theta)\left|1_{a_i}\right\rangle\left|1_{b_i}\right\rangle\right)$$

Since $R_y(\alpha)R_y(\beta)=R_y(\alpha+\beta)$ and $R_y(0)=I$ ( $I$ is the quantum unitary transformation), $R_y(-(k_i\oplus k_{(i+1)\bmod n}+1+r_i)\theta)R_y((k_i\oplus k_{(i+1)\bmod n}+1+r_i)\theta)=R_y(0)=I$ , i.e.

$$R_y(-(k_i\oplus k_{(i+1)\bmod n}+1+r_i)\theta)\left|\phi_i''\right\rangle=\frac{1}{\sqrt{2}}\left(\left|0_{a_i}0_{b_i}\right\rangle+\left|1_{a_i}1_{b_i}\right\rangle\right)=\left|\phi_i\right\rangle$$

If $k_i=1$, it can be proved similarly for

$$R_y(-(k_i\oplus k_{(i+2)\bmod n}+1+r_i)\theta)\left|\phi_i''\right\rangle=\frac{1}{\sqrt{2}}\left(\left|0_{a_i}0_{b_i}\right\rangle+\left|1_{a_i}1_{b_i}\right\rangle\right)=\left|\phi_i\right\rangle$$

And the particles $a_i$ and $b_i$ are in the quantum entangled state $\left|\phi_i\right\rangle=\frac{1}{\sqrt{2}}\left(\left|0_{a_i}0_{b_i}\right\rangle+\left|1_{a_i}1_{b_i}\right\rangle\right)$.

Furthermore, in *Step* 7, Alice measures $a_i$ and $b_i$ in the Bell basis, which can only obtain $c_i = 0$ ($i = 0,1,\cdots,n-1$). Thus, Alice passes Bob's identity authentication when $c_0 = c_1 = \cdots = c_{i-1} = 0$.

It is obvious that Bob can authenticate Alice's identity in the same manner, meaning the new protocol can be extended to a bi-directional authentication protocol.

## 3 Analyze of the new protocol

Suppose Eve is an attacker.

### 3.1. Impersonates attack

**A. Alice to Bob communication: Eve impersonates Bob**

(1) Eve intercepts the particles that Alice sends to Bob and imitates Bob by performing transformations on the particles. Since Eve has no information about the shared key *K*, Eve can only guess the rotation angle $(k_i'+1)\theta$ where $i = 0,1,\cdots,n-1$ and $k_i' = 0$ or 1. After applying $R_y((k_i'+1)\theta)$ on the particle $a_i$, the entangled state is

$$\frac{1}{\sqrt{2}}\left(R_y((k_i'+1+r_i)\theta)\left|0_{a_i}\right\rangle\left|0_{b_i}\right\rangle + R_y((k_i'+1+r_i)\theta)\left|1_{a_i}\right\rangle\left|1_{b_i}\right\rangle\right)$$

Then $(a_0, a_1, \cdots, a_{n-1})$ is send back to Alice. And $R_y(-(k_i \oplus k_{(i+1)\bmod n} + 1 + r_i)\theta)$ or $R_y(-(k_i \oplus k_{(i+2)\bmod n} + 1 + r_i)\theta)$ will be applied on the particle $a_i$. Without loss of generality, let the transformation applying on the particle $a_i$ be $R_y(-(k_i \oplus k_{(i+1)\bmod n} + 1 + r_i)\theta)$. And the entangled state is $\frac{1}{\sqrt{2}}\left(R_y((k_i' - k_i \oplus k_{(i+1)\bmod n})\theta)\left|0_{a_i}\right\rangle\left|0_{b_i}\right\rangle + R_y((k_i' - k_i \oplus k_{(i+1)\bmod n})\theta)\left|1_{a_i}\right\rangle\left|1_{b_i}\right\rangle\right)$.

If $k_i' = k_i \oplus k_{(i+1)\bmod n}$, i.e., Eve guesses correctly, the entangled state is $\frac{1}{\sqrt{2}}\left(\left|0_{a_i}\right\rangle\left|0_{b_i}\right\rangle + \left|1_{a_i}\right\rangle\left|1_{b_i}\right\rangle\right)$, and this bit can be passed by Alice.

If $k_i' \neq k_i \oplus k_{(i+1)\bmod n}$, i.e., Eve guesses incorrectly and $k_i' - k_i \oplus k_{(i+1)\bmod n} = -1$ or 1, the entangled state is $\frac{\cos((k_i' - k_i \oplus k_{(i+1)\bmod n})\theta)/2}{\sqrt{2}}\left(\left|0_{a_i}\right\rangle\left|0_{b_i}\right\rangle + \left|1_{a_i}\right\rangle\left|1_{b_i}\right\rangle\right) - \frac{\sin((k_i' - k_i \oplus k_{(i+1)\bmod n})\theta)/2}{\sqrt{2}}\left(\left|0_{a_i}\right\rangle\left|1_{b_i}\right\rangle - \left|1_{a_i}\right\rangle\left|0_{b_i}\right\rangle\right)$, and this bit can be passed by Alice with probability $\cos^2\theta/2$.

In summary, the *i*-th bit can be passed by Alice with probability $\frac{1+\cos^2\theta/2}{2}$. And *n* bit can be passed with probability $(\frac{1+\cos^2\theta/2}{2})^n$, i.e., the probability of Eve successfully impersonating Bob is $(\frac{1+\cos^2\theta/2}{2})^n$. And the success probability will be sufficiently small when $n$ is large.

(2) Eve impersonates Bob by directly sending particle $a_i$ back to Alice. After Alice receives particle $a_i$, the entangled state is

$$|\phi_i'\rangle = \frac{1}{\sqrt{2}}\left(R_y\left(r_i\theta\right)|0_a\rangle|0_b\rangle + R_y\left(r_i\theta\right)|1_a\rangle|1_b\rangle\right)$$

And $R_y(-(k_i \oplus k_{(i+1)\bmod n} + 1 + r_i)\theta)$ or $R_y(-(k_i \oplus k_{(i+2)\bmod n} + 1 + r_i)\theta)$ will be applied on the particle $a_i$. Without loss of generality, let the transformation applying on the particle $a_i$ be $R_y(-(k_i \oplus k_{(i+1)\bmod n} + 1 + r_i)\theta)$.

$$\frac{1}{\sqrt{2}}\left(R_y(-(k_i \oplus k_{(i+1)\bmod n} + 1)\theta)|0_a\rangle|0_b\rangle + R_y(-(k_i \oplus k_{(i+1)\bmod n} + 1)\theta)|1_a\rangle|1_b\rangle\right),\ k_i = 0$$

If $k_i \oplus k_{(i+1)\bmod n} + 1 = 1$, the entangled state is $\frac{\cos(-\theta/2)}{\sqrt{2}}\left(|0_{a_i}\rangle|0_{b_i}\rangle + |1_{a_i}\rangle|1_{b_i}\rangle\right) - \frac{\sin(-\theta/2)}{\sqrt{2}}\left(|0_{a_i}\rangle|1_{b_i}\rangle - |1_{a_i}\rangle|0_{b_i}\rangle\right)$, and this bit can be passed by Alice with probability $\cos^2\theta/2$.

If $k_i \oplus k_{(i+1)\bmod n} + 1 = 2$, this bit can be passed by Alice with probability $\cos^2\theta$.

In summary, the *i*-th bit can be passed by Alice with probability $\frac{\cos^2\theta + \cos^2(\theta/2)}{2}$. The probability of Eve successfully impersonating Bob is $(\frac{\cos^2\theta + \cos^2(\theta/2)}{2})^n$. And the success probability will be also sufficiently small when $n$ is large.

**B. Bob to Alice communication: Eve impersonates Alice**

Eve impersonates Alice by sending quantum states to Bob. Since Bob does not have the ability to authenticate Alice's identity—meaning he performs transformations on all incoming quantum states, Eve can directly send $n$ particles in the $|0\rangle$ state to Bob and obtain information about the shared key from the returned states.

Bob receives $n$ particles and $R_y((k_i \oplus k_{(i+1)\bmod n} + 1)\theta)$ or $R_y((k_i \oplus k_{(i+2)\bmod n} + 1)\theta)$ will be applied on each particle. Without loss of generality, let the rotation angle be $k'\theta$ where $k' = 1\ or\ 2$. After the transformation, the state is

$$R_y(k'\theta)|0\rangle = \cos\frac{k'\theta}{2}|0\rangle + \sin\frac{k'\theta}{2}|1\rangle$$

$$= \begin{cases} \cos\frac{\pi}{8}|0\rangle + \sin\frac{\pi}{8}|1\rangle, k' = 1 \\ \cos\frac{\pi}{4}|0\rangle + \sin\frac{\pi}{4}|1\rangle, k' = 2 \end{cases}$$

And Eve can measure this state in the diagonal basis and obtains $|-\rangle$, $k' = 1$. However, the probabilities of $k' = k_i \oplus k_{(i+1)\bmod n} + 1$ and $k' = k_i \oplus k_{(i+2)\bmod n} + 1$ are both $1/2$, i.e.,

Eve can't know whether $k_i \oplus k_{(i+1)\bmod n} = 0$ or $k_i \oplus k_{(i+2)\bmod n} = 0$. Of course, Eve can't obtain any shared information. The main reason is that the authentication information is generated based on pre-shared information, which increases the randomness of the secret information relative to Eve. And the new protocol can resist impersonates attack.

### 3.2. CNOT attack

In the Perturbation phase, Alice needs to send particle $a_i$ ($i = 0,1,\cdots,n-1$) to Bob, which is entangled with particle $b_i$, and they are in the entangled state $|\phi_i'\rangle = \frac{1}{\sqrt{2}}\left(R_y(r_i\theta)|0_{a_i}\rangle|0_{b_i}\rangle + R_y(r_i\theta)|1_{a_i}\rangle|1_{b_i}\rangle\right)$. However, $b_i$ is not sent to Bob. Therefore, if Eve implements a CNOT attack, Eve can only perform quantum transformations on particle $a_i$. According to Ref.[40], generally, $a_i$ is used as the control qubit. Eve also needs to prepare an ancillary qubit $c_i$, that is, to perform a CNOT operation on $c_i$ and $a_i$. After the CNOT operation, the quantum state $|\delta_i\rangle$ of $c_i$, $a_i$ and $b_i$ is related to the initial state of $c_i$.

(i) When the initial state of $c_i$ is $|0_{c_i}\rangle$,

$$|\delta_i\rangle = \frac{1}{\sqrt{2}}\left(|0_{c_i}\rangle R_y(r_i\theta)|0_{a_i}\rangle|0_{b_i}\rangle + |1_{c_i}\rangle R_y(r_i\theta)|1_{a_i}\rangle|1_{b_i}\rangle\right)$$

Eve can only measure $c_i$ and $a_i$.

(1) If $c_i$ is measured individually, then the quantum state of $a_i$ and $b_i$ is as follows:

① When the measurement result is 0, $a_i$ and $b_i$ are in the quantum state $R_y(r_i\theta)|0_{a_i}\rangle|0_{b_i}\rangle$. After the encryption and decryption phases, $a_i$ and $b_i$ are in the quantum state $|0_{a_i}\rangle|0_{b_i}\rangle$. At this point, performing a Bell-state measurement yields probabilities of $\frac{1}{2}$ for both 0 and 1. In the absence of a CNOT attack, the measurement result would definitely be 0. Therefore, Alice detects that the protocol has been attacked with probability $1-\frac{1}{2^n}$.

② When the measurement result is 1, $a_i$ and $b_i$ are in the quantum state $R_y(r_i\theta)|1_{a_i}\rangle|1_{b_i}\rangle$. As in case ①, Alice also detects that the protocol has been attacked with probability $1-\frac{1}{2^n}$.

(2) If $c_i$ and $a_i$ are measured, the entanglement between $a_i$ and $b_i$ must be destroyed.

① When $r_i = 0$, the probabilities of $a_i$ and $b_i$ being in the quantum state

$\left|0_{a_i}\right\rangle\left|0_{b_i}\right\rangle$ and $\left|1_{a_i}\right\rangle\left|1_{b_i}\right\rangle$ are both $\frac{1}{2}$. After the encryption and decryption phases, $a_i$ and $b_i$ are in the quantum state $\left|0_{a_i}\right\rangle\left|0_{b_i}\right\rangle$ or $\left|1_{a_i}\right\rangle\left|1_{b_i}\right\rangle$. Alice detects that the protocol has been attacked with probability $1-\frac{1}{2^n}$.

② When $r_i=1$, the probabilities of $a_i$ and $b_i$ being in the quantum state $\left|0_{a_i}\right\rangle\left|0_{b_i}\right\rangle$, $\left|0_{a_i}\right\rangle\left|1_{b_i}\right\rangle$, $\left|1_{a_i}\right\rangle\left|0_{b_i}\right\rangle$ and $\left|1_{a_i}\right\rangle\left|1_{b_i}\right\rangle$ are $\frac{1}{2}\cos^2\frac{\pi}{8}$, $\frac{1}{2}\sin^2\frac{\pi}{8}$, $\frac{1}{2}\sin^2\frac{\pi}{8}$ and $\frac{1}{2}\cos^2\frac{\pi}{8}$ respectively. After the encryption and decryption phases, $a_i$ and $b_i$ are in the quantum state $R_y(-\theta)\left|0_{a_i}\right\rangle\left|0_{b_i}\right\rangle$, $R_y(-\theta)\left|0_{a_i}\right\rangle\left|1_{b_i}\right\rangle$, $R_y(-\theta)\left|1_{a_i}\right\rangle\left|0_{b_i}\right\rangle$ and $R_y(-\theta)\left|1_{a_i}\right\rangle\left|1_{b_i}\right\rangle$. At this point, performing a Bell-state measurement yields probabilities of $\frac{1}{2}(\cos^4\frac{\pi}{8}+\sin^4\frac{\pi}{8})=\frac{3}{8}$ for 0. Alice detects that the protocol has been attacked with probability $1-(\frac{3}{8})^n$.

Furthermore, if Eve does not perform a measurement at this point, then after the encryption and decryption phases, the quantum state of $c_i$, $a_i$ and $b_i$ is $\frac{1}{\sqrt{2}}\left(\left|0_{c_i}\right\rangle\left|0_{a_i}\right\rangle\left|0_{b_i}\right\rangle+\left|1_{c_i}\right\rangle\left|1_{a_i}\right\rangle\left|1_{b_i}\right\rangle\right)$. At this time, performing a Bell-state measurement on $a_i$ and $b_i$ yields probabilities of 0.5 for both 0 and 1. That is, Alice detects that the protocol has been attacked with probability $1-\frac{1}{2^n}$.

In summary, the protocol can resist the CNOT attack when $n$ is large.

(ii) Likewise, when the initial state of $c_i$ is $\left|1_{c_i}\right\rangle$, $\frac{1}{\sqrt{2}}\left(\left|0_{c_i}\right\rangle+\left|1_{c_i}\right\rangle\right)$ or $\frac{1}{\sqrt{2}}\left(\left|0_{c_i}\right\rangle-\left|1_{c_i}\right\rangle\right)$, the protocol can also resist the CNOT attack.

### 3.3. Entanglement measurement attack

In the entanglement measurement attack, the core is to construct a quantum transformation $U_F$, which acts on a two-qubit state, yielding the following quantum state:

$$U_F(|0\rangle|\varepsilon\rangle)=a_0|0\rangle|0\rangle+b_0|0\rangle|1\rangle+c_0|0\rangle|+\rangle+d_0|0\rangle|-\rangle$$
$$U_F(|1\rangle|\varepsilon\rangle)=a_1|1\rangle|0\rangle+b_1|1\rangle|1\rangle+c_1|1\rangle|+\rangle+d_1|1\rangle|-\rangle$$
$$U_F(|+\rangle|\varepsilon\rangle)=a_+|+\rangle|0\rangle+b_+|+\rangle|1\rangle+c_+|+\rangle|+\rangle+d_+|+\rangle|-\rangle$$
$$U_F(|-\rangle|\varepsilon\rangle)=a_-|-\rangle|0\rangle+b_-|-\rangle|1\rangle+c_-|-\rangle|+\rangle+d_-|-\rangle|-\rangle$$

Where $|\varepsilon\rangle$ is the attacker's auxiliary qubit. Through $U_F$, the qubit sent by the sender is entangled with $|\varepsilon\rangle$. However, the receiver does not know that the qubit sent by the sender is entangled with $|\varepsilon\rangle$. After the receiver performs a quantum measurement, the attacker can perform a quantum measurement based on the information fed back by the receiver to the sender, thereby obtaining the secret information.

If the quantum state sent by the sender is $\alpha|0\rangle+\beta|1\rangle$, and $U_F$ acts on this qubit and the auxiliary qubit, the resulting quantum state is

$$\begin{aligned}U_F((\alpha|0\rangle+\beta|1\rangle)|\varepsilon\rangle) &= \alpha(a_0|0\rangle|0\rangle+b_0|0\rangle|1\rangle+c_0|0\rangle|+\rangle+d_0|0\rangle|-\rangle)\\ &\quad+\beta(a_1|1\rangle|0\rangle+b_1|1\rangle|1\rangle+c_1|1\rangle|+\rangle+d_1|1\rangle|-\rangle)\end{aligned}$$

The quantum state is then sent to the receiver. After the receiver measures the first qubit, the resulting quantum state is $a_0|0\rangle+b_0|1\rangle+c_0|+\rangle+d_0|-\rangle)$ or $a_1|0\rangle+b_1|1\rangle+c_1|+\rangle+d_1|-\rangle)$. Obviously, the attacker cannot obtain any information about $\alpha$ or $\beta$ through measurement, meaning that the entanglement measure attack can't obtain any information about the quantum probability amplitude. In the new protocol, the secret information is encoded in the quantum probability amplitude. Thus, the new protocol can resist the entanglement measure attack.

### 3.4. Backward security

In this protocol, Bob performs a rotation transformation on $a_i$ according to the shared key, thereby obtaining the quantum state $|\phi_i''\rangle$. The authentication information to be sent to Alice is encoded in the probability amplitudes of $|\phi_i''\rangle$. Since $a_i$ is entangled with $b_i$, and the attacker cannot control $b_i$, Eve can, at most, only obtain the form of the quantum state in which $a_i$ resides during the perturbation and encryption phase. Furthermore, during the perturbation phase, Alice performs $R_y(r_i\theta)$ on $a_i$ according to the random sequence $r=(r_0,r_1,\cdots,r_{n-1})$, thereby adding a perturbation probability amplitude to the quantum state in which $a_i$ resides. Therefore, the attacker cannot accurately obtain the secret information encoded in $|\phi_i''\rangle$, meaning that they cannot determine the value of $k_i\oplus k_{(i+1)\bmod n}$ or $k_i\oplus k_{(i+2)\bmod n}$. Consequently, even if Eve obtains the specific form of the quantum state in which $a_i$ resides at the current moment, it will not have any impact on the security of the subsequent identity authentication process. That is, the protocol has backward security.

### 3.5. Comparison

In the protocol by Chen et al.,[32] Alice and Bob share a *n*-bit strings $k_{ide}=(k_{ide}^1,\cdots,k_{ide}^n)$, where $k_{ide}^i$ is *n* bits, i.e., $n^2$-bit information. And the new protocol only needs share *n*-bit information, representing a $1/n$ reduction compared to Chen et al.'s protocol.

In the protocol by Dutta et al., [33] Alice and Bob share a 2*n*-bit information

$K=\left(k_1,k_2,\cdots,k_{2n}\right)$. Both parties respectively prepare an *n*-bit decoy sequence and an *n*-bit authentication sequence based on these 2*n* bits of information. However, in our protocol, Alice generates the authentication sequence based on a set of random sequences, and Bob selects $(k_i\oplus k_{(i+1)\bmod n})\theta$ or $(k_i\oplus k_{(i+2)\bmod n})\theta$ as the authentication information based on the value of the shared information $k_i$. Compared to Dutta et al.'s protocol, where each 2 bits of shared information can only prepare 1 bit of quantum state, our protocol improves the utilization efficiency of the shared information. Bob only needs to prepare *n* bits of authentication information based on the shared *n*-bit information to complete the authentication. Under the same security conditions, the amount of shared information is reduced by half.

In the protocol by Dutta et al., if the attacker measures and obtains $|+\rangle$ in the horizontal basis, the authentication code sent is $|0\rangle$, which obviously has the risk of leaking secret information. However, in the new protocol, the classical information shared is encoded in the probability amplitudes of the quantum state, and the form of the quantum state is

$$\frac{1}{\sqrt{2}}\left(R_y((k_i\oplus k_{(i+1)\bmod n}+r_i)\theta)\left|0_{a_i}\right\rangle\left|0_{b_i}\right\rangle+R_y((k_i\oplus k_{(i+1)\bmod n}+r)\theta)\left|1_{a_i}\right\rangle\left|1_{b_i}\right\rangle\right)$$

or

$$\frac{1}{\sqrt{2}}\left(R_y((k_i\oplus k_{(i+2)\bmod n}+r_i)\theta)\left|0_{a_i}\right\rangle\left|0_{b_i}\right\rangle+R_y((k_i\oplus k_{(i+2)\bmod n}+r_i)\theta)\left|1_{a_i}\right\rangle\left|1_{b_i}\right\rangle\right).$$

This quantum state is transmitted over the public channel. Regardless of the measurement method used by the attacker, they cannot exactly know the probability amplitudes of the quantum state. At the same time, a random perturbation probability amplitude $R_y(r_i\theta)$ is added to the probability amplitudes of the quantum state, and $r_i$ is randomly generated by Alice before identity authentication. Therefore, the new scheme ensures that there is no risk of leakage for each bit under special measurements, and thus there is no security risk.

Furthermore, Bob can authenticate Alice's identity in the same way, meaning that the new protocol can be extended to a bi-directional quantum identity authentication protocol.

## 4 Conclusion

In this paper, the quantum identity authentication via perturbed probability amplitude is presented. The protocol has forward security, and can resist impersonates attack, CNOT attack and entanglement measurement attack. And each bit of the secret message has no risk of leakage under special measurements. Furthermore, it can be extended to a bidirectional quantum identity authentication protocol. Future work may provide additional security analyses.

## Acknowledgements

This work is supported by National Key Research and Development Program of China (Grant No. 2020YFA0309702), National Natural Science Foundation of China (Grant No. 61502526), NSAF (Grant No. U2130205), Natural Science Foundation of Henan (Grant No. 202300410532).